\documentclass{article}

\usepackage{arxiv}

\usepackage[utf8]{inputenc} 
\usepackage[T1]{fontenc}    
\usepackage{hyperref}       
\usepackage{url}            
\usepackage{booktabs}       
\usepackage{amsfonts}       
\usepackage{nicefrac}       
\usepackage{microtype}      
\usepackage{lipsum}
\usepackage{graphicx}
\graphicspath{ {./images/} }

\usepackage{graphicx,verbatim}
\usepackage{amsmath}
\usepackage{amssymb}
\usepackage{algorithm}
\usepackage{algpseudocode}
\usepackage{booktabs} 
\usepackage{multirow}

\title{BiM-GeoAttn-Net: Linear-Time Depth Modeling with Geometry-Aware Attention for 3D Aortic Dissection CTA Segmentation}

\author{
 Yuan Zhang \\
  College of Computer Science\\
  Sichuan Normal University\\
  Chengdu, Sichuan 610068, China \\
  \texttt{wq986788@gmail.com} \\
   \And
 Lei Liu \\
  Zhejiang University \& Ant Group \\
  Hangzhou, Zhejiang, China \\
  \texttt{liulei1497@gmail.com} \\
  \And
 Jialin Zhang \\
  State Key Laboratory of Public Big Data\\
  Guizhou University\\
  Guiyang, Guizhou 550025, China \\
  \texttt{gs.jlzhang25@gzu.edu.cn} \\
  \And
 Ya-Nan Zhang \\
  College of Computer Science\\
  Sichuan Normal University\\
  Chengdu, Sichuan 610068, China \\
  \texttt{zyn962464@gmail.com} \\
  \And
  Ling Wang \\
  College of Computer Science\\
  Sichuan Normal University\\
  Chengdu, Sichuan 610068, China \\
  \texttt{lingwang@sicnu.edu.cn} \\
  \And
 Nan Mu $^{*}$ \\
  College of Computer Science\\
  Sichuan Normal University\\
  Chengdu, Sichuan 610068, China \\
  \texttt{nanmu@sicnu.edu.cn} \\
}

\begin{document}
\maketitle
\begin{abstract}
Accurate segmentation of aortic dissection (AD) lumens in CT angiography (CTA) is essential for quantitative morphological assessment and clinical decision-making. However, reliable 3D delineation remains challenging due to limited long-range context modeling, which compromises inter-slice coherence, and insufficient structural discrimination under low-contrast conditions. To address these limitations, we propose \textbf{BiM-GeoAttn-Net}, a lightweight framework that integrates linear-time depth-wise state-space modeling with geometry-aware vessel refinement. Our approach is featured by \textbf{Bidirectional Depth Mamba (BiM)} to efficiently capture cross-slice dependencies and \textbf{Geometry-Aware Vessel Attention (GeoAttn)} module that employs orientation-sensitive anisotropic filtering to refine tubular structures and sharpen ambiguous boundaries. Extensive experiments on a multi-source AD CTA dataset demonstrate that BiM-GeoAttn-Net achieves a Dice score of 93.35\% and an HD95 of 12.36 mm, outperforming representative CNN-, Transformer-, and SSM-based baselines in overlap metrics while maintaining competitive boundary accuracy. These results suggest that coupling linear-time depth modeling with geometry-aware refinement provides an effective, computationally efficient solution for robust 3D AD segmentation.
\end{abstract}


\section{Introduction}

\begin{figure*}[ht]
    \centering
    \includegraphics[width=1\textwidth]{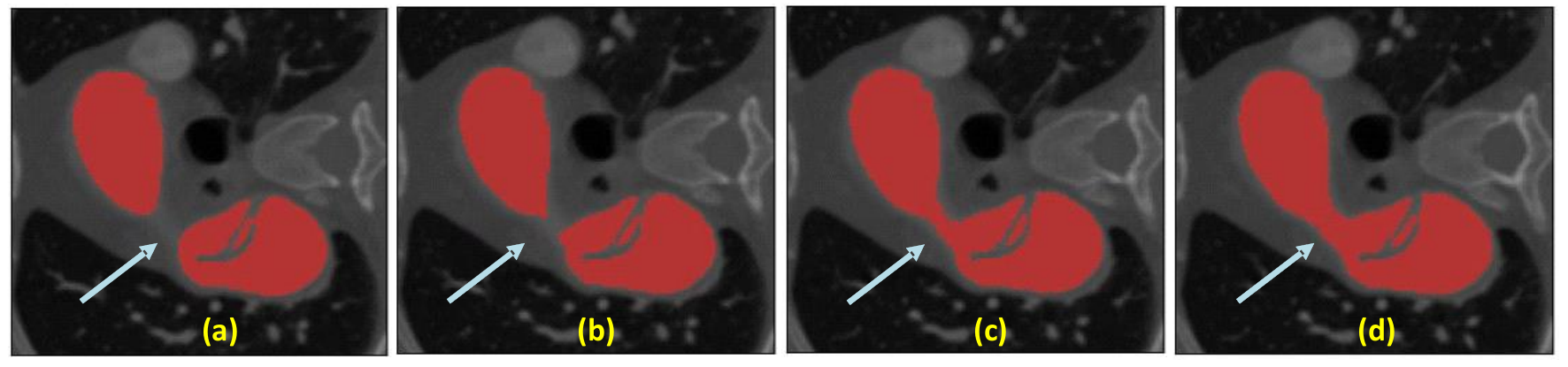}
    \caption{Challenges in AD CTA segmentation. Four consecutive axial slices exhibit strong cross-slice continuity, subtle boundary contrast, and morphological variation. Ground-truth masks are in red.}
    \label{fig:f1}
\end{figure*}

Accurate segmentation of aortic dissection (AD) lumens is critical for morphological assessment and longitudinal monitoring of Type-B AD~\cite{Li2018CascadedAD,Cao2019TBADDL}, enabling cohort-level analysis via standardized volumetric descriptors~\cite{Chen2021MultiStageAD}. However, robust segmentation in CT angiography (CTA) is hindered by the aorta's complex, tortuous geometry and substantial appearance variations along the centerline. Anisotropic voxel spacing further compromises inter-slice coherence, while boundaries remain subtle against surrounding tissues. As shown in Fig.~\ref{fig:f1}, although the dissected lumen requires depth-axis continuity, slice-wise variations and boundary ambiguity frequently result in fragmented predictions and unstable delineations.

Conventional 3D CNNs effectively capture local texture and boundary cues~\cite{mu2023attention}, yet their limited receptive fields hinder long-range dependency modeling. Transformer-based architectures improve global context aggregation via self-attention~\cite{hu2024slimmable}, but full 3D attention is computationally expensive, and practical approximations may compromise boundary precision. Recently, state-space models (SSMs) such as Mamba offer linear-time long-sequence modeling~\cite{Ma2024UMamba}. However, existing SSM-based medical segmentation methods typically adopt generic sequence construction without explicitly aligning modeling direction with anatomical continuity, and purely global modeling remains insufficient for low-contrast boundary refinement without structural priors~\cite{Xie2025VesselMamba}.

 
To address these limitations, we propose BiM-GeoAttn-Net, a bottleneck enhancement framework that integrates efficient depth-axis state-space modeling with geometry-aware vessel refinement. Specifically, a Bidirectional Depth Mamba (BiM) module performs linear-time modeling along the depth axis to strengthen cross-slice consistency. Built upon BiM, a Geometry-Aware Vessel Attention (GeoAttn) module incorporates orientation-sensitive anisotropic filtering with dual attention mechanisms to emphasize tubular structures and sharpen ambiguous boundaries. The cascaded BiM-GeoAttn design directly addresses two dominant failure modes in AD CTA segmentation: depth-wise discontinuity and boundary degradation under low contrast.

The main contributions are summarized as follows:

(1) We propose \textbf{BiM-GeoAttn-Net}, a lightweight 3D segmentation framework tailored for aortic dissection lumens, addressing challenges in long-range context modeling and low-contrast boundary delineation.

(2) We design a \textbf{Bidirectional Depth Mamba} module for linear-time cross-slice coherence and a \textbf{Geometry-Aware Vessel Attention} module to employ orientation-sensitive anisotropic filtering for enhancing tubular structures.


(3) Extensive experiments demonstrate consistent improvements over representative CNN-, Transformer-, and SSM-based baselines in overlap accuracy, while maintaining competitive boundary precision and computational efficiency.

\section{Methodology}

\subsection{Motivation and Overview}

\begin{figure*}[t]
    \centering
    \includegraphics[width=1\linewidth]{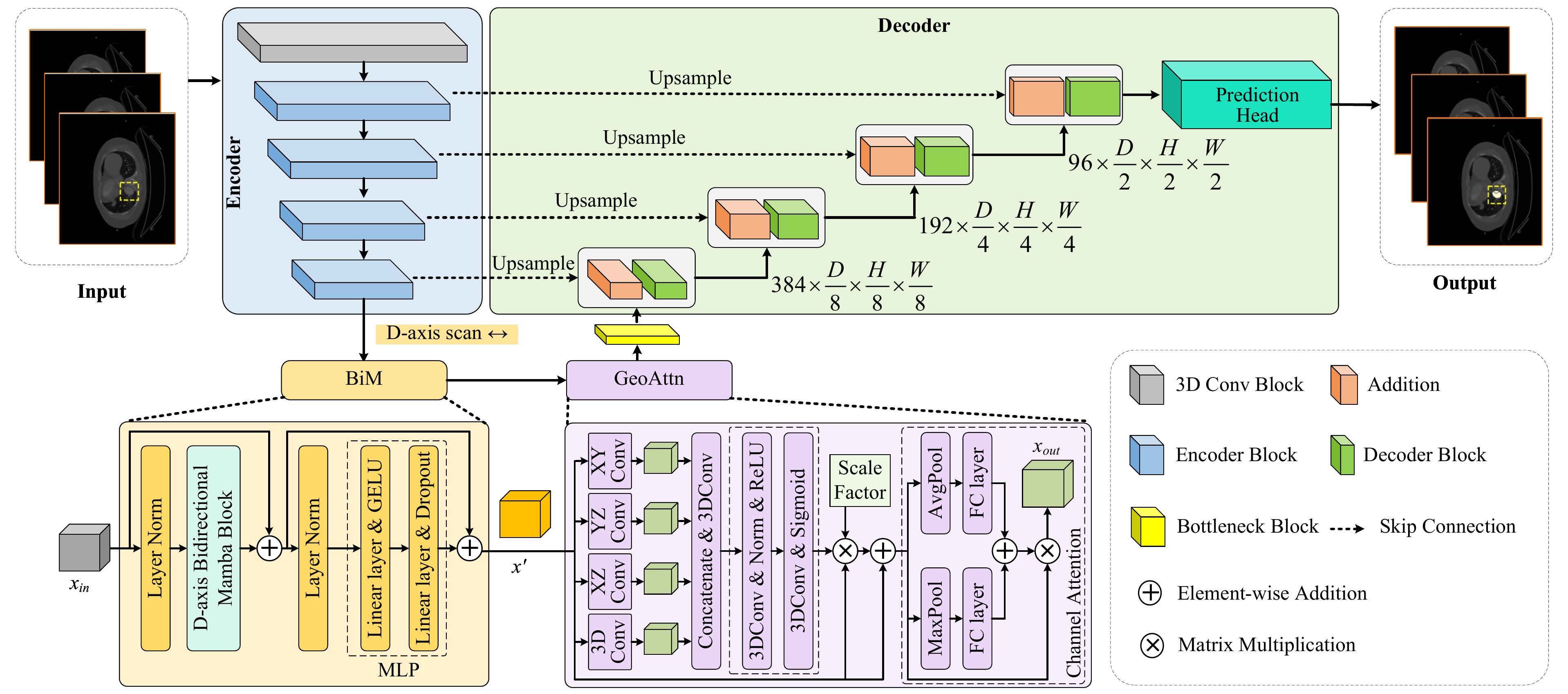}
    \caption{Overall architecture of the proposed BiM-GeoAttn-Net.}
    \label{fig:f2}
\end{figure*}

As illustrated in Fig.~\ref{fig:f2}, we build upon the 3D nnU-Net~\cite{Isensee2023nnUNetv2}, which adopts a U-shaped encoder–decoder architecture with multi-scale feature extraction and skip connections. Although nnU-Net provides strong volumetric representation capability, AD segmentation remains challenged by two key issues: (i) inter-slice inconsistency in anisotropic CTA volumes, leading to fragmented lumen predictions, and (ii) boundary ambiguity under low contrast, which degrades thin-structure delineation.

To explicitly address these limitations, we introduce BiM-GeoAttn-Net by augmenting the bottleneck of nnU-Net with two complementary modules. A Bidirectional Depth Mamba (BiM) module performs depth-axis state-space modeling to enhance long-range cross-slice coherence with near-linear complexity. Built upon this global modeling stage, a Geometry-Aware Vessel Attention (GeoAttn) module incorporates vessel-structure priors via orientation-sensitive filtering and dual attention mechanisms, refining tubular boundaries and suppressing background interference.

The cascaded BiM–GeoAttn design unifies depth-wise continuity modeling and local structure-aware refinement within a lightweight framework. The enhanced features are residually fused and forwarded to the decoder for final lumen segmentation, with training supervised by a hybrid objective to mitigate foreground–background imbalance.

\subsection{Bidirectional Depth Mamba}

The BiM module models long-range dependencies along the depth axis without incurring the quadratic cost of full 3D self-attention. This is particularly important for maintaining coherent lumen predictions across adjacent slices.

Let $\boldsymbol{x}_{in}\in\mathbb{R}^{B\times C\times D\times H\times W}$ denote the input feature map. We reshape spatial locations into depth-wise sequences
\[
\widetilde{\boldsymbol{x}}=\mathrm{Reshape}(\boldsymbol{x}_{in})
\in\mathbb{R}^{(BHW)\times D\times C},
\]
where each spatial position forms a 1D sequence along $D$. A Mamba state-space model $\mathcal{M}(\cdot)$ is applied bidirectionally:
\begin{equation}
\boldsymbol{z}^{\rightarrow}=\mathcal{M}(\widetilde{\boldsymbol{x}}), \quad
\boldsymbol{z}^{\leftarrow}=\mathrm{Rev}\!\left(\mathcal{M}\!\left(\mathrm{Rev}(\widetilde{\boldsymbol{x}})\right)\right),
\end{equation}
\begin{equation}
\boldsymbol{z}=\boldsymbol{z}^{\rightarrow}+\boldsymbol{z}^{\leftarrow}.
\end{equation}
The fused representation is reshaped back and added residually:
\begin{equation}
\boldsymbol{x}_{1}=\boldsymbol{x}_{in}+\mathrm{Reshape}^{-1}(\boldsymbol{z}).
\end{equation}
Then a lightweight channel-mixing MLP with pre-normalization further refines the features:
\begin{equation}
\boldsymbol{x}'=\boldsymbol{x}_{1}+\mathrm{MLP}(\mathrm{Norm}(\boldsymbol{x}_{1})),
\end{equation}
where $\mathrm{MLP}(\boldsymbol{u})=W_2\,\sigma(W_1\boldsymbol{u})$ with expansion ratio $r$.

Based on SSM-based scanning, BiM achieves near-linear complexity with respect to $D$ while aggregating context from both directions, thereby stabilizing inter-slice continuity and boundary prediction.

\subsection{Geometry-Aware Vessel Attention}

The GeoAttn module refines vessel representations by combining direction-aware filtering with spatial and channel attention. Given an input feature map 
$\boldsymbol{X}\in\mathbb{R}^{B\times C\times D\times H\times W}$, it outputs $\boldsymbol{Y}$ of the same shape.

To capture orientation-sensitive tubular structures, we employ three plane-aligned anisotropic convolutions together with a standard 3D convolution:
\begin{equation}
\begin{aligned}
\boldsymbol{F}_{xy} &= \mathrm{Conv}_{1\times3\times3}(\boldsymbol{X}), \\
\boldsymbol{F}_{yz} &= \mathrm{Conv}_{3\times1\times3}(\boldsymbol{X}), \\
\boldsymbol{F}_{xz} &= \mathrm{Conv}_{3\times3\times1}(\boldsymbol{X}), \\
\boldsymbol{F}_{3d} &= \mathrm{Conv}_{3\times3\times3}(\boldsymbol{X}),
\end{aligned}
\end{equation}
where each branch produces $C_b$ channels. The multi-branch features are concatenated and fused via a $1\times1\times1$ convolution:
\begin{equation}
\boldsymbol{F}=
\mathrm{Conv}_{1\times1\times1}
\big[
\boldsymbol{F}_{xy},
\boldsymbol{F}_{yz},
\boldsymbol{F}_{xz},
\boldsymbol{F}_{3d}
\big].
\end{equation}
Then, a lightweight bottleneck generates a spatial attention map
\begin{equation}
\boldsymbol{A}=
\sigma\!\left(
\mathrm{Conv}_{1\times1\times1}
\left(
\delta\!\left(
\mathrm{IN}
\left(
\mathrm{Conv}_{1\times1\times1}(\boldsymbol{F})
\right)
\right)
\right)
\right),
\end{equation}
which modulates the input through a residual gate:
\begin{equation}
\boldsymbol{X}_s=
\boldsymbol{X}\odot (1+\gamma\boldsymbol{A}).
\end{equation}
Finally, channel attention is applied using global average and max pooling:
\begin{equation}
\boldsymbol{W}=
\sigma\!\left(
\mathrm{MLP}(\mathrm{GAP}(\boldsymbol{X}_s))+
\mathrm{MLP}(\mathrm{GMP}(\boldsymbol{X}_s))
\right),
\end{equation}
\begin{equation}
\boldsymbol{Y}=
\boldsymbol{X}_s\odot \boldsymbol{W}.
\end{equation}

By combining plane-aligned anisotropic filtering with dual attention mechanisms, GeoAttn enhances geometry-consistent vessel regions and sharpens ambiguous boundaries. Positioned after BiM, it complements depth-wise coherence modeling with orientation-aware boundary refinement.

\subsection{Loss Function}

We adopt a hybrid Dice–Cross Entropy objective to address foreground–background imbalance. Let $p_{i,c}$ denote the predicted probability of voxel $i$ belonging to class $c$, and $g_{i,c}\in\{0,1\}$ the corresponding ground truth indicator. The total loss is
\begin{equation}
\mathcal{L}_{\text{total}} 
= \lambda_{\text{dice}}\mathcal{L}_{\text{Dice}} 
+ \lambda_{\text{ce}}\mathcal{L}_{\text{CE}}.
\end{equation}
The soft Dice loss averaged over $K$ classes is
\begin{equation}
\mathcal{L}_{\text{Dice}}
= 1 - \frac{1}{K}\sum_{c=1}^{K}
\frac{2\sum_{i} p_{i,c} g_{i,c} + \epsilon}
{\sum_{i} p_{i,c}^{2} + \sum_{i} g_{i,c}^{2} + \epsilon},
\end{equation}
where $\epsilon$ ensures numerical stability. The cross-entropy term is computed voxel-wise:
\begin{equation}
\mathcal{L}_{\text{CE}}
= -\frac{1}{N}\sum_{i}\sum_{c} g_{i,c}\log(p_{i,c}).
\end{equation}
We set $\lambda_{\text{dice}}=\lambda_{\text{ce}}=1$ to balance global shape alignment and voxel-level accuracy.

\section{Experiments}

This Section describes the experimental setup, quantitative evaluation, and ablation studies used to assess the effectiveness and robustness of our method.

\subsection{Dataset}

We train and evaluate our model on a curated dataset for Stanford Type-B Aortic Dissection (TBAD) vessel segmentation, termed \textit{Dataset500\_VesselTBAD}. The dataset combines cases from two publicly available sources: 
(1) \textbf{ImageTBAD}~\cite{Abaid2024TBAD2D}: This dataset provides expert-annotated 3D CTA scans with labels for the true lumen (TL), false lumen (FL), and false lumen thrombosis (FLT). We selected 32 cases based on annotation quality and pathological diversity.
(2) \textbf{TBD-CTA}~\cite{Mayer2024FigshareAD}: A large-scale TBAD CTA dataset covering varying degrees of thrombosis and dissection extent. We included 39 cases with complete TL/FL annotations.

In total, 71 cases are split at the patient level into training, validation, and test sets with $N_{\text{train}}=50$, $N_{\text{val}}=7$, and $N_{\text{test}}=14$. Each subset contains samples from both sources: ImageTBAD (22/3/7) and TBD-CTA (28/4/7) for train/val/test, respectively. The patient-level split ensures that scans from the same subject do not appear across different subsets.

We formulate the task as binary segmentation, extracting the vessel foreground (TL+FL) from the background. FLT annotations are not treated as an independent class in this study.

\subsection{Implementation Details and Evaluation Metrics}

Our method was implemented in PyTorch within the nnU-Net framework and trained on NVIDIA RTX 3090 GPUs with CUDA support. Automatic mixed-precision (AMP) training~\cite{Micikevicius2017MixedPrecision} was adopted to improve computational efficiency and memory usage. We followed the nnU-Net 3d\_fullres configuration and employed a patch-based training scheme with randomly cropped $128 \times 128 \times 128$ patches and a batch size of 2. The network was trained for 400 epochs using SGD with a cosine annealing learning rate schedule. Weighted deep supervision was applied at multiple resolution levels, and the final model was selected based on the best validation performance.

Segmentation performance was evaluated using Dice Similarity Coefficient (Dice), Intersection over Union (IoU), Recall, Precision, and 95\% Hausdorff Distance (HD95). Dice, IoU, Recall, and Precision measure volumetric overlap accuracy ($\uparrow$), while HD95 assesses boundary agreement ($\downarrow$). All metrics were computed on the test set and averaged across cases.

\subsection{Comparison with State-of-the-Art Methods}

We compare BiM-GeoAttn-Net with representative 3D segmentation models spanning CNN-, Transformer-, and SSM-based paradigms, including Attention U-Net~\cite{Oktay2018AttentionUNet}, nnU-Net v2~\cite{Isensee2023nnUNetv2}, Swin-Unet~\cite{Hu2022SwinUNet}, SegFormer3D~\cite{Li2022SegFormer3D}, and Mamba-UNet~\cite{Zhang2024MambaUNet}. All methods are trained and evaluated under identical settings for fair comparison.

\begin{table*}[t]
\centering
\caption{Quantitative comparison with state-of-the-art methods on the AD dataset. 
Best results are shown in bold. ($\uparrow$ higher is better; $\downarrow$ lower is better.)}
\label{tab:t1}
\small
\setlength{\tabcolsep}{4pt}
\renewcommand{\arraystretch}{1.15}
\resizebox{\textwidth}{!}{%
\begin{tabular}{l|ccccc|cc}
\hline
\multirow{2}{*}{Models} 
& \multicolumn{5}{c|}{Segmentation Performance} 
& \multicolumn{2}{c}{Training Cost} \\
\cline{2-8}
& Prec. $\uparrow$ 
& Rec. $\uparrow$ 
& Dice $\uparrow$ 
& IoU $\uparrow$ 
& HD95 (mm) $\downarrow$
& Time (min/epoch) $\downarrow$ 
& Memory (GB) $\downarrow$ \\
\hline
Attention U-Net  & 89.71 & 90.76 & 90.22 & 82.26 & \textbf{10.77} & 3.2 & 8.3 \\
nnU-Net          & 90.46 & 91.49 & 90.84 & 83.64 & 18.51 & 2.8 & 7.6 \\
Swin-UNet        & 87.78 & 91.73 & 89.62 & 81.62 & 20.01 & 4.1 & 10.7 \\
SegFormer3D      & 87.53 & 88.12 & 88.43 & 79.30 & 24.15 & 4.6 & 11.2 \\
Mamba-UNet       & 88.83 & 91.52 & 89.43 & 83.12 & 12.20 & 3.5 & 8.9 \\
\hline
\textbf{BiM-GeoAttn-Net} 
& \textbf{92.04} 
& \textbf{94.85} 
& \textbf{93.35} 
& \textbf{87.53} 
& 12.36 
& 2.9 
& 8.0 \\
\hline
\end{tabular}%
}
\end{table*}

\textbf{Quantitative Results.} As shown in Table~\ref{tab:t1}, our method achieves the best overall performance across overlap metrics. It attains the highest Dice (93.35\%), IoU (87.53\%), Recall (94.85\%), and Precision (92.04\%). Compared with nnU-Net, Dice and IoU improve by 2.51\% and 3.89\%, respectively. 
Regarding boundary accuracy, Attention U-Net achieves the lowest HD95 (10.77\,mm), while our model (12.36\,mm) remains competitive and substantially outperforms nnU-Net (18.51\,mm) and SegFormer3D (24.15\,mm). These results indicate that our approach improves volumetric overlap while maintaining strong boundary fidelity.

\textbf{Computational Efficiency.} As reported in Table~\ref{tab:t1}, our model requires 2.9\,min/epoch and 8.0\,GB memory, which is comparable to nnU-Net (2.8\,min, 7.6\,GB). In contrast, Transformer-based models such as SegFormer3D and Swin-Unet incur higher computational cost. This demonstrates that the proposed method achieves improved accuracy without introducing substantial overhead.

\begin{figure*}[tb]
    \centering
    \includegraphics[width=\linewidth]{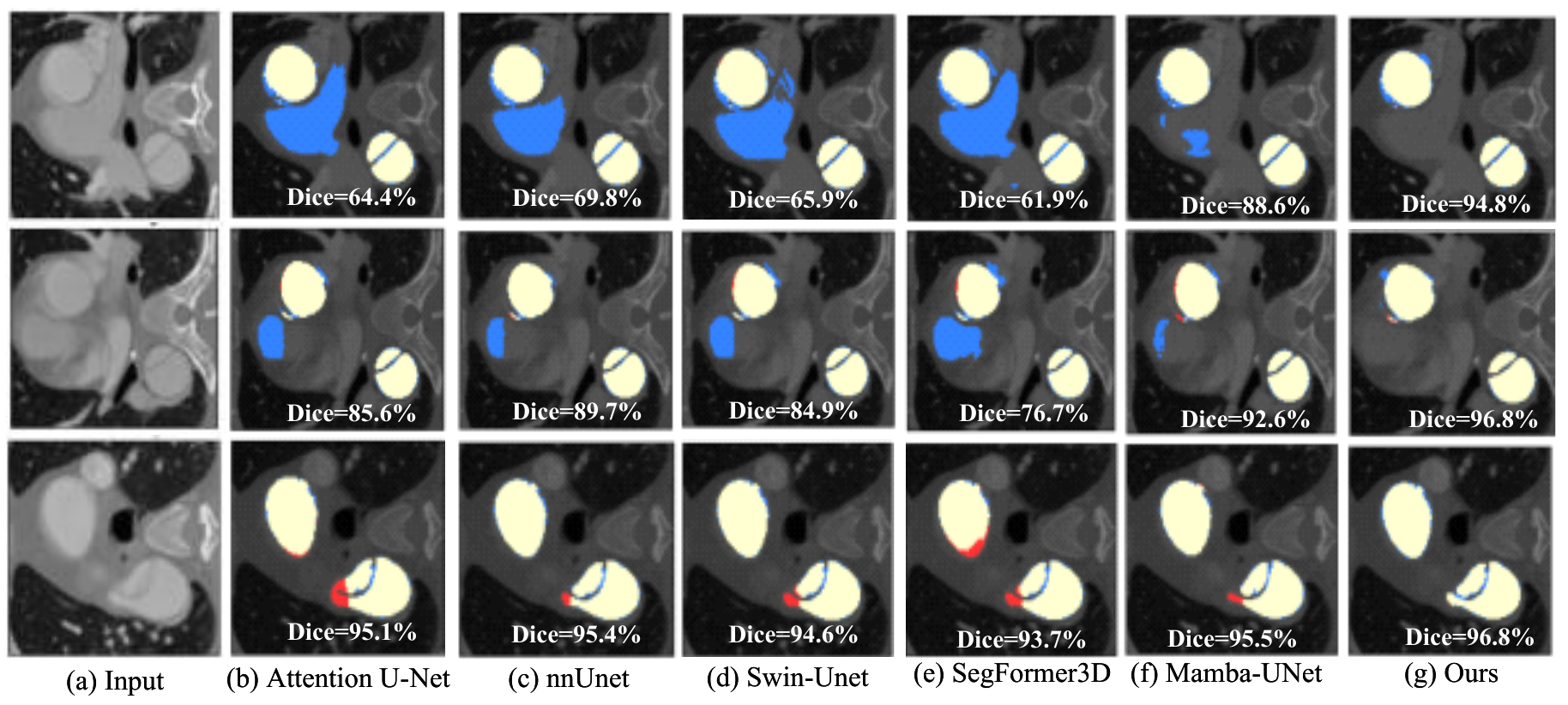}
    \caption{Qualitative comparison on representative AD test cases. Yellow indicates correct overlap, blue denotes false positives, and red denotes false negatives. Compared with CNN-, Transformer-, and SSM-based baselines, BiM-GeoAttn-Net achieves sharper boundaries and improved depth-wise continuity.}    
    \label{fig:f3}
\end{figure*}

\textbf{Qualitative Results.}  Fig.~\ref{fig:f3} presents representative visual comparisons. Transformer-based methods tend to produce over-segmentation around ambiguous boundaries, while nnU-Net and Mamba-UNet occasionally miss fine vascular structures in low-contrast regions. In contrast, our model yields fewer false positives and false negatives, resulting in more consistent lumen continuity and sharper boundary delineation. These observations align with the quantitative improvements in both overlap and boundary metrics.

\subsection{Ablation Study}

To assess the contribution of each component, we evaluate four configurations: (1) baseline (B), (2) B+BiM, (3) B+GeoAttn, and (4) B+BiM+GeoAttn (\textit{i.e.}, BiM-GeoAttn-Net). Quantitative results are summarized in Table~\ref{tab:t2}.

\begin{table}[htbp]
\centering
\caption{Ablation results on the AD dataset. Best results are shown in bold. ($\uparrow$ higher is better; $\downarrow$ lower is better.)}
\label{tab:t2}
\setlength{\tabcolsep}{6pt}
\renewcommand{\arraystretch}{1.2}
\fontsize{12pt}{14pt}\selectfont
\resizebox{\linewidth}{!}{
\begin{tabular}{lccccc}
\toprule
Models & Dice (\%) $\uparrow$ & IoU (\%) $\uparrow$ & Recall (\%) $\uparrow$ & Precision (\%) $\uparrow$ & HD95 (mm) $\downarrow$ \\
\midrule
B              & 89.77 & 81.60 & 92.11 & 87.73 & 29.08 \\
B+BiM           & 91.12 & 83.69 & 94.41 & 89.65 & 20.35 \\
B+GeoAttn       & 90.67 & 83.25 & 93.17 & 88.46 & 25.48 \\
B+BiM+GeoAttn   & \textbf{93.35} & \textbf{87.53} & \textbf{94.85} & \textbf{92.04} & \textbf{12.36} \\
\bottomrule
\end{tabular}}
\end{table}

\textbf{Quantitative Analysis.} The baseline achieves a Dice of 89.77\% with relatively high HD95 (29.08\,mm), indicating limited boundary precision. Incorporating BiM improves Dice to 91.12\% and reduces HD95 to 20.35\,mm, reflecting enhanced cross-slice continuity. Adding GeoAttn alone yields moderate gains (Dice 90.67\%) and improved Recall, suggesting better local structure sensitivity. 
Combining both modules delivers the best performance, with Dice reaching 93.35\% and HD95 decreasing substantially to 12.36\,mm. These results demonstrate the complementarity of the two components: BiM strengthens depth-wise global coherence, while GeoAttn refines directional vessel boundaries.

\begin{figure}[t]
    \centering
    \includegraphics[width=\linewidth]{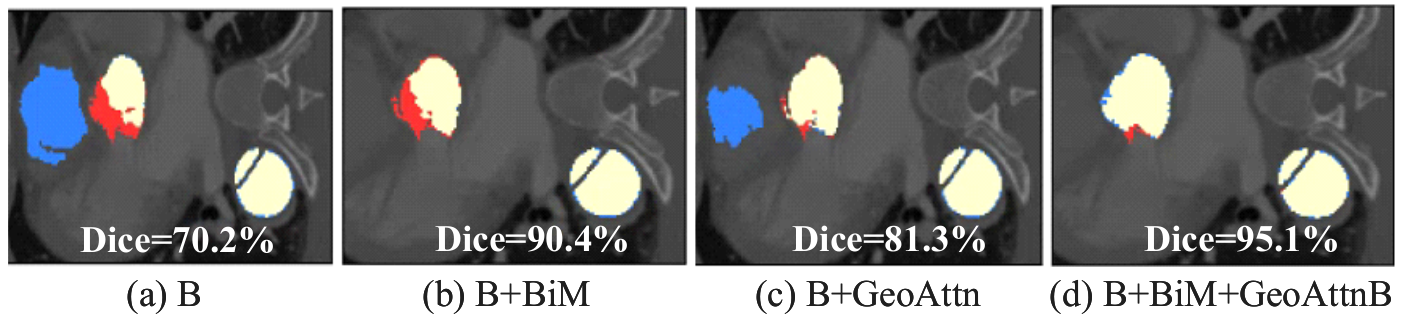}
    \caption{Qualitative ablation comparison. Yellow indicates correct overlap, blue false positives, and red false negatives. The combined model yields the most consistent continuity and boundary delineation.}
    \label{fig:f4}
\end{figure}

\textbf{Qualitative Analysis.} Fig.~\ref{fig:f4} further illustrates the effect of each module. The baseline exhibits leakage and fragmented predictions. BiM reduces false positives and improves cross-slice consistency, whereas GeoAttn enhances boundary adherence but may not fully suppress large-scale interference. The combined model achieves the most accurate delineation, with fewer missed regions and improved structural continuity, corroborating the quantitative findings.

\section{Conclusions}

We presented BiM-GeoAttn-Net, a segmentation framework for aortic dissection (AD) CTA that integrates depth-wise long-range modeling with geometry-aware vessel refinement. By combining a Bidirectional Depth Mamba (BiM) module and a Geometry-Aware Vessel Attention (GeoAttn) module at the network bottleneck, the proposed method addresses slice-wise discontinuity and low-contrast boundary ambiguity in AD segmentation.
Extensive experiments and ablation studies demonstrate consistent improvements over representative 3D baselines, yielding enhanced volumetric coherence, sharper boundary delineation, and improved robustness in challenging regions. These findings suggest that coupling linear-time depth modeling with geometry-aware refinement provides an effective strategy for accurate and stable 3D AD CTA segmentation, with potential to support quantitative lumen analysis and computer-aided clinical assessment.

\section*{Acknowledgments}
This work was supported by the National Natural
Science Foundation of China (62006165) and the Natural Science Foundation of Sichuan Province (2025ZNSFSC1477).

\bibliographystyle{unsrt}  
\bibliography{references}

@inproceedings{Li2018CascadedAD,
  author    = {Li, X. and others},
  title     = {Lumen Segmentation of Aortic Dissection with Cascaded Convolutional Network},
  booktitle = {Proceedings of STACOM/MICCAI},
  year      = {2018},
  publisher = {Springer-Verlag},
  series    = {Lecture Notes in Computer Science}
}

@article{Cao2019TBADDL,
  author    = {Cao, Y. and others},
  title     = {Fully Automatic Segmentation of Type {B} Aortic Dissection from {CTA} Images Enabled by Deep Learning},
  journal   = {European Journal of Radiology},
  year      = {2019},
  publisher = {Elsevier}
}

@article{Chen2021MultiStageAD,
  author    = {Chen, X. and others},
  title     = {Multi-Stage Learning for Segmentation of Aortic Dissections Using a Prior Aortic Anatomy Simplification},
  journal   = {Medical Image Analysis},
  year      = {2021},
  publisher = {Elsevier}
}

@article{Abaid2024TBAD2D,
  title     = {Exploratory analysis of Type B Aortic Dissection (TBAD) segmentation in 2D CTA images using various kernels},
  author    = {Abaid, Ayman and Ilancheran, Srinivas and Iqbal, Talha and Hynes, Niamh and Ullah, Ihsan},
  journal   = {Computerized Medical Imaging and Graphics},
  volume    = {118},
  pages     = {102460},
  year      = {2024},
  publisher = {Elsevier},
  doi       = {10.1016/j.compmedimag.2024.102460},
  url       = {https://doi.org/10.1016/j.compmedimag.2024.102460}
}

@misc{Ma2024UMamba,
  author       = {Ma, X. and others},
  title        = {U-Mamba: Enhancing Long-Range Dependency for Biomedical Image Segmentation},
  year         = {2024},
  howpublished = {arXiv preprint},
  note         = {arXiv preprint (identifier to be completed)}
}

@article{Xie2025VesselMamba,
  author    = {Xie, Y. and others},
  title     = {VesselMamba: {3D} Vessel Segmentation in {CTA} Using Mamba with Enhanced Spatial-Channel Attention},
  journal   = {Biomedical Signal Processing and Control},
  year      = {2025},
  publisher = {Elsevier}
}

@misc{Mayer2024FigshareAD,
  title        = {{Aortic Dissection Dataset and Segmentations}},
  author       = {Mayer, C. and others},
  year         = {2024},
  howpublished = {Figshare},
  url          = {https://figshare.com/articles/dataset/Aortic_Dissection_Dataset_and_Segmentations/22269091}
}

@article{Micikevicius2017MixedPrecision,
  title   = {Mixed precision training},
  author  = {Micikevicius, Paulius and Narang, Sharan and Alben, Jonah and Diamos, Gregory and Elsen, Erich and Garcia, David and Ginsburg, Boris and Houston, Michael and Kuchaiev, Oleksii and Venkatesh, Ganesh and others},
  journal = {arXiv preprint arXiv:1710.03740},
  year    = {2017},
  url     = {https://arxiv.org/abs/1710.03740}
}

@article{Hu2022SwinUNet,
  title     = {Swin-Unet: Unet-like Pure Transformer for Medical Image Segmentation},
  author    = {Hu, Cao and Gu, Yu and Zhang, Jiang and Luo, Han and Liu, Li and Wang, Yaming and Tian, Qi},
  journal   = {IEEE Transactions on Medical Imaging},
  volume    = {41},
  number    = {8},
  pages     = {2078--2091},
  year      = {2022},
  publisher = {IEEE},
  doi       = {10.1109/TMI.2022.3156452}
}

@article{Isensee2023nnUNetv2,
  title     = {nnU-Net v2: Evolution of a self-configuring framework for medical image segmentation},
  author    = {Isensee, Fabian and Jaeger, Paul F and Kohl, Simon A and Petersen, Jens and Maier-Hein, Klaus H},
  journal   = {Medical Image Analysis},
  volume    = {80},
  pages     = {102683},
  year      = {2023},
  publisher = {Elsevier},
  doi       = {10.1016/j.media.2023.102683}
}

@article{Li2022SegFormer3D,
  title   = {SegFormer3D: Simple and Efficient 3D Semantic Segmentation with Transformers},
  author  = {Li, Xiang and Wang, Wenhai and Hu, Xiaolin and Yang, Jian and Liu, Dengxin and Zhou, Yukun and Lu, Lewei and Luo, Ping},
  journal = {arXiv preprint arXiv:2207.06275},
  year    = {2022}
}

@article{Oktay2018AttentionUNet,
  title   = {Attention U-Net: Learning where to look for the pancreas},
  author  = {Oktay, Ozan and Schlemper, Jo and Folgoc, Loic Le and Lee, Matthew and Heinrich, Mattias and Misawa, Kazunari and Mori, Kensaku and McDonagh, Steven and Hammerla, Nils Y and Kainz, Bernhard and others},
  journal = {arXiv preprint arXiv:1804.03999},
  year    = {2018}
}

@article{Zhang2024MambaUNet,
  title     = {Mamba-UNet: Mamba-based Universal Network for Medical Image Segmentation},
  author    = {Zhang, Yixiao and Chen, Yukun and Li, Jia and Wang, Ziyi and Zhang, Hao},
  journal   = {Computer Methods and Programs in Biomedicine},
  volume    = {240},
  pages     = {108421},
  year      = {2024},
  publisher = {Elsevier},
  doi       = {10.1016/j.cmpb.2024.108421}
}

@article{mu2023attention,
  title     = {An attention residual u-net with differential preprocessing and geometric postprocessing: Learning how to segment vasculature including intracranial aneurysms},
  author    = {Mu, Nan and Lyu, Zonghan and Rezaeitaleshmahalleh, Mostafa and Tang, Jinshan and Jiang, Jingfeng},
  journal   = {Medical Image Analysis},
  volume    = {84},
  pages     = {102697},
  year      = {2023},
  publisher = {Elsevier}
}

@article{hu2024slimmable,
  title     = {Slimmable transformer with hybrid axial-attention for medical image segmentation},
  author    = {Hu, Yiyue and Mu, Nan and Liu, Lei and Zhang, Lei and Jiang, Jingfeng and Li, Xiaoning},
  journal   = {Computers in Biology and Medicine},
  volume    = {173},
  pages     = {108370},
  year      = {2024},
  publisher = {Elsevier}
}






\end{document}